\documentstyle[preprint,eqsecnum,aps]{revtex}

\begin{document}
\draft
\preprint{HEP/123-qed}
\title{
Difference between the cleaved surface and the polished surface on the 
Raman spectra of La$_{2-x}$Sr$_x$CuO$_4$
}

\author{S. Sugai, Y. Takayanagi, and N. Hayamizu}
\address{Department of Physics, Faculty of Science, Nagoya University, 
Chikusa-ku, Nagoya 464-8602, Japan} 

\date{\today}
\maketitle

\begin{abstract}
The Raman spectra on the cleaved surface of La$_{2-x}$Sr$_x$CuO$_4$ 
is found to be different from those on the mechanically polished or 
chemically etched surface.  
On the cleaved surface, the step-like increase of the scattering 
intensity below 700 cm$^{-1}$ is washed away and the two-phonon 
peaks are preserved clearly even at the optimum doping.  
The difference is demonstrated on the optimally doped crystal at 
$x=0.15$ and the stripe phase crystal at $x=0.115$.  
The stripe phase is stabilized by defects induced by mechanical 
polishing.
\end{abstract}


\narrowtext

The surface treatment is very important for Raman scattering in high 
$T_{\rm c}$ superconductors, because visible light penetrates into 
the sample by only several hundreds \AA \ from the surface. 
Up to now the most experiments in La$_{2-x}$Sr$_x$CuO$_4$ (LSCO) 
have been made on the surfaces prepared 
mechanical polishing or followed by chemical etching with a 
bromine-ethanol solution \cite{1,2,3,4,5,6,7,8} except for the 
experiments of antiferromagnetic insulator La$_2$CuO$_4$ by 
Lyons {\it et al}. \cite{9,10} who used as-grown or cleaved surfaces.  
In the case of Bi$_2$Sr$_2$CaCu$_2$O$_{8+\delta}$ (Bi2212), cleaved 
surfaces have been usually used, because the cleavage is very easy 
\cite{11,12,13,14}.  
It is known that the cleaved surface is the ideal surface, but 
it is difficult to obtain a good cleaved surface in 
LSCO or YBa$_2$Cu$_3$O$_{7-\delta}$ (YBCO).  
We executed Raman scattering on the cleaved surfaces of LSCO 
selected from a large number of broken flakes of singles crystals.  
Then we found that the spectra are very different from those of 
polished or chemically etched surfaces.  

We prepared the single crystals at the best condition as far as we 
could.  
The single crystals were synthesized by the traveling-solvent 
floating-zone (TSFZ) method in the infrared-radiation-heating furnace 
with four oval millers (Crystal system, FZ-T-4000).  
The melted portion is supported by the surface tension without using a 
crucible.  
A ceramic disk (the thickness is 3-4 mm) of LSCO with excess 
CuO$_2$ solvent (the ratio among La$_2$O$_3$, CuO, and SrCO$_3$ is the 
same as the initial composition of the flax method \cite{15}) is 
inserted between a single crystal seed and a LSCO ceramic source rod.  
The solvent disk is heated by focused light and the melted portion is 
moved at the speed of 1 mm/h.  
The typical sizes of the single crystal is 4 mm$\phi \times$50 mm length.  

This synthesis method ensures the sample free from the contamination of 
the crucible and the superconducting transition temperature 
($T_{\rm c}$) is the same as that of the ceramic.  
Usually single crystals synthesized by the flax method at which the 
mixture of LSCO and CuO$_2$ solvent is cooled down slowly in a crucible 
includes about 0.5 \% or more amount of Al or Pt, which reduces the 
$T_{\rm c}$ to about 37 K at highest from 42 K of the ceramic LSCO 
($x=0.15$).  
Moreover the single crystal synthesized by the TSFZ method has uniform 
Sr concentration, because the Sr concentration in the liquid is kept 
constant during the crystal growth by solving the ceramic source with 
the same Sr concentration.  
On the other hand the single crystal synthesized by the flax method has 
different Sr concentration from the center to the surface of the crystal, 
because the Sr concentration in the liquid changes along the phase 
boundary between the liquid and the solid as temperature decreases.  
The quality of the crystal was estimated by the x-ray diffraction, the 
temperature dependence of the resistivity along the $a$ (or $b$) axis 
and the $c$ axis, and the polarized microscope image.  
There is no impurity phase in the x-ray diffraction.  
The temperature dependent resistivity is consistent with the reported 
results for both in-plane and out-of-plane directions \cite{16}. 
The $T_{\rm c}$ determined from the mid-point of the transition of the 
resistivity in the CuO$_2$ plane is 42 K for LSCO ($x=0.15$) and 33 K 
for LSCO ($x=0.115$).  
The temperature width between the onset and the end of the 
superconducting transition is 1.5 K for LSCO ($x=0.15$) and 2.5 K for 
LSCO ($x=0.115$).  
The cross polarized microscope image of the sample surface shows that 
the whole area of the sample is composed of the (001) plane.  
Small holes with a diameter of about a few tens $\mu$m were included 
in some part of the crystal.  
We measured the Raman spectra in the area free from the holes by 
monitoring the irradiated sample image through TV camera set in the 
spectrometer.  

The crystal axes were determined by the x-ray Raue pattern.  
The most part of the crystal rod was cut into many cubes with the 
sizes $3 \times 3 \times 3$ mm the surfaces of which were perpendicular 
to the crystal axes.  
The cubes were cleaved into several tens of pieces.  
A few pieces with the best condition were used for the Raman scattering 
measurement.  
For each Sr concentration, a piece of $3 \times 3 \times 1$ mm was cut 
and polished by diamond abrasive films successively decreasing the 
particle diameter from 3, 0.5, to 0.1 $\mu$m or from 3, 1, to 0.3 
$\mu$m in the clean bench.  
We used n-decane as polishing liquid and hexane as rinse liquid.  
Hexane gives the better result than methanol or ethanol as for the 
contamination on the surface, because alcohol easily absorbs moisture.  
The obtained sample was set in the cryostat and the ambient gas was 
exchanged to helium within 10 minutes and then the sample was cooled 
down to 5 K very slowly taking more than 12 hours.  

	Raman spectra were measured in a quasi-back scattering 
configuration utilizing a triple monochromator (JASCO, NR-1810), 
a liquid nitrogen cooled CCD detector (Princeton, 1100PB), and a 
5145 \AA \ Ar-ion laser (Spectra Physics, stabilite 2017).  
The laser beam of 10 mW was focused on the area of 50$\times$500 
$\mu$m$^2$.  
The increase of temperature by the laser beam irradiation was less 
than 2 K at 5 K.  
The same spectra were measured four times to remove the cosmic ray 
noise by comparing the intensities at each channel.  
The wide energy spectra covering $12-7000$ cm$^{-1}$ was obtained 
by connecting 17 spectra with narrow energy ranges after correcting 
the spectroscopic efficiency of the optical system.  
The same spot on the surface was measured during the temperature 
variation by correcting the sample position which was monitored by 
a TV camera inside the spectrometer.  

Figure 1 shows polarized Raman spectra of LSCO ($x=0.15$) at 5 K 
and 300 K on the cleaved surface and the polished surface.  
The final polishing particle diameter is 0.3 $\mu$m.  
The polarization configuration $(xy)$ denotes that the electric 
field of the incident light is parallel to the $x$ axis and the 
scattered light with the electric field parallel to the y axis is 
detected.  
The $a=[100]$ and $b=[010]$ directions are parallel to the Cu-O-Cu 
bonds and the $x$ and $y$ directions are parallel to [110] and 
[1\=10].  
The $(xy)$ spectrum includes the $B_{\rm 1g}$ symmetry modes, the 
$(ab)$ spectrum $B_{\rm 2g}$, the $(xx)$ spectrum 
$A_{\rm 1g}$+$B_{\rm 2g}$, and the $(aa)$ spectrum 
$A_{\rm 1g}$+$B_{\rm 1g}$.  
The $A_{\rm 1g}$ spectrum is obtained by 
$0.5\times($(xx)$+(aa)-$(xy)$-(ab))$.  
The overall spectra are the same for both cleavage and polishing 
as shown in Fig. 1.  
However, the superconducting gap structure (discussed later) in all 
symmetry spectra, many two-phonon peaks from 750 cm$^{-1}$ to 
1400 cm$^{-1}$ in the $A_{\rm 1g}$ spectrum observed on the cleaved 
surface at 5 K is missing in the spectra on the polished surface.  
Instead, the step-like structure emerges below 700 cm$^{-1}$ on the 
polished surface.  

	Figure 2 shows the low energy spectra.  
The intensity of two-phonon peaks at 750-1400 cm$^{-1}$ decreases 
in the $A_{\rm 1g}$ spectrum, when the surface is polished.   
Instead, the broad phonon peak appears at 580 cm$^{-1}$ and the 
scattering intensity shows a step-like increase below 700 cm$^{-1}$ 
independently of the symmetry.  
The height of the step decreases to about a half, when the surface 
is polished by the abrasive film with the diamond diameter of 0.1 
$\mu$m.  
Therefore this structure is attributed to the defect-induced extrinsic 
phonon peaks.  
This step-like structure cannot be removed, even if the surface is 
etched chemically \cite{7,8}.  
It indicates that the chemical etching leaves contaminations.  
The width of the single-phonon peak below about 700 cm$^{-1}$ is 
much broader on the polished surface than on the cleaved surface.  

The superconducting gap structure can be used for the estimation 
of the surface and the sample quality.  
Figure 3 shows the $B_{\rm 1g}$, $B_{\rm 2g}$, and $A_{\rm 1g}$ 
spectra at 5K and 40 K and the differential spectra between these 
temperatures.  
The $B_{\rm 1g}$ electronic Raman spectrum represents the 
electronic excitation around the $(00)-(\pi 0)$ direction where 
the $d$-wave superconducting gap is at the maximum \cite{12,17}.  
The $B_{\rm 2g}$ spectrum represents the excitation around the 
$(00)-(\pi \pi)$ direction where the gap is at the node.  
The superconducting gap structure is very clearly observed on the 
cleaved surface.  
The peak energies are 204 cm$^{-1}$ for $B_{\rm 1g}$ and 112 
cm$^{-1}$ for $B_{\rm 2g}$, which are consistent with the results 
on the chemically etched surface \cite{4,5}.  
The gap structure disappeared completely on the polished surface.  

Figure 4 shows the Raman spectra of LSCO ($x=0.115$) at 5 K and 
300 K on the cleaved surface.  
The diameter of the final polishing particles is 0.1 $\mu$m.  
The $T_{\rm c}$ decreases 
at $x=0.115$ by a few degree from the expected $T_{\rm c}$ on the 
$T_{\rm c}$ vs $x$ curve \cite{18}.  
This is known as the "1/8 problem".  
In the case of La$_{2-x}$Ba$_x$CuO$_4$ and 
La$_{2-x-y}$Nd$_y$Sr$_x$CuO$_4$, the $T_{\rm c}$ is completely 
suppressed \cite{18,19,20,21}.  
Tranquada {\it et al}. \cite{22,23} proposed the stripe structure model in 
which the antiferromagnetic spin stripes are separated by the 
charge domain walls.  
The two-magnon Raman peak at 2373 cm$^{-1}$ at 300 K splits into 
double peaks at 1910 cm$^{-1}$ and 3037 cm$^{-1}$ at 5 K, as 
the stripe structure develops at low temperatures.  
The split of the two-magnon peaks is caused by the different 
magnetic excitation energy in the two-magnon Raman process 
near the charge domain walls and in the spin strips.  
The details will be published elsewhere \cite{24}.  
The incommensurate split of the magnetic spot at $(\pi\pi)$ has 
been observed by neutron scattering in the whole range of $x$ in 
LSCO \cite{25,26,27,28,29,30,31}.  
In Raman scattering, however, the split of the two-magnon peak 
is observed at very narrow Sr concentration region the width 
of which is only about 0.01 around $x=0.115$ in 
consistent with the decrease of $T_{\rm c}$ obtained by the 
electric conductivity and magnetic susceptibility measurements \cite{18}.  
The narrow Sr concentration makes the observation of the split 
two-magnon peak difficult.  
The crystal synthesized by the TSFZ method has the advantage of 
the uniform Sr concentration over the crystal made by the slow 
cooling method.  
At 300K split two-magnon peaks decreases in intensity and the 
single two-magnon peak is left in the $B_{\rm 1g}$ spectrum of the 
cleaved surface, but split two-magnon humps are still noticeable 
in the case of the polished surface.  
It indicates that the stripe structure is almost vanished on the 
cleaved surface at 300 K, but is left on the polished surface.   
The scattering intensity is enhanced, as the stripe structure 
develops upon decreasing temperature.  
The strong enhancement of the scattering intensity on the 
polished surface also indicates that the defects stabilize the 
stripe structure.  
The polishing induces the extra phonon peak around 580 cm$^{-1}$ 
instead of the decrease of the intensity of the two-phonon peaks.  
The superconducting gap peak is observed in the $B_{\rm 2g}$ 
spectra on the cleaved surface, but not observed on the polished 
surface.  

In conclusion the polished surface loses the superconducting 
gap structure and the two-phonon scattering intensity.  
On the other hand the extrinsic phonon peaks appear around 580 
cm$^{-1}$ and the scattering intensity increases below it.  
It causes the defect-induced step-like structure at 700 cm$^{-1}$ 
in the spectra.  
The gap structure is recovered by the chemical etching, but the 
phonon spectrum is retained in the similar structure as on the 
polished surface.  
It suggests the importance of the measurement on the cleaved 
surface, although it needs large single crystals.  
The stripe structure at $x=0.115$ is stabilized by the defects 
induced by polishing.

Acknowledgments.  
This work was supported by CREST of the Japan Science and 
Technology Corporation.

\begin{figure}
\caption[]{
(gray) Raman spectra of La$_{1.85}$Sr$_{0.15}$CuO$_4$ on the cleaved 
and the polished surfaces at 5 K and 300 K.  
The intensity is divided by $n(\omega,T)+1$, so as to be expressed by 
the Raman susceptibility, where $n(\omega, T)$ is the Bose-Einstein 
factor.  
}
\label{fig1}
\end{figure}

\begin{figure}
\caption[]{
Comparison of the low energy spectra at 5 K between on the cleaved 
surface and the polished surface of La$_{1.85}$Sr$_{0.15}$CuO$_4$.  
}
\label{fig2}
\end{figure}

\begin{figure}
\caption[]{
Superconducting gap structure on the cleaved surface of 
La$_{1.85}$Sr$_{0.15}$CuO$_4$.  
}
\label{fig3}
\end{figure}

\begin{figure}
\caption[]{
(gray) Raman spectra in the stripe phase of 
La$_{1.885}$Sr$_{0.115}$CuO$_4$.  
}
\label{fig4}
\end{figure}

\end{document}